\documentclass[preprint,12pt]{elsarticle}

\usepackage{graphicx}
\usepackage{amssymb}
\usepackage{color}

\def\ket#1{ \left\vert  #1  \right\rangle }
\def\bra#1{ \left\langle  #1 \right\vert}
\def\spr#1#2{ \left\langle #1 \left\vert \right. #2 \right\rangle }
\def\etal{\textit{et al.}}

\def\cor#1{{\color{black}{#1}}}

\journal{Physics Letters A}

\begin{document}

\begin{frontmatter}

\title{Fidelity estimation between two finite ensembles
        \\ of unknown pure equatorial qubit states}

\author{Michael Siomau}
\ead{siomau@physi.uni-heidelberg.de}

\address[ger]{Physikalisches Institut, Heidelberg Universit\"{a}t,
D-69120 Heidelberg, Germany}
\address[bel]{\cor{Department of Theoretical Physics,
Belarussian State University, 220030 Minsk, Belarus}}

\begin{abstract}
Suppose, we are given two finite ensembles of pure qubit states, so
that the qubits in each ensemble are prepared in identical (but
unknown for us) states lying on the equator of the Bloch sphere.
What is the best strategy to estimate fidelity between these two
finite ensembles of qubit states? We discuss three possible
strategies for the fidelity estimation. We show that the best
strategy includes two stages: a specific unitary transformation on
two ensembles and state estimation of the output states of this
transformation.
\end{abstract}
\end{frontmatter}

The state of a quantum system can be perfectly reconstructed only by
computing statistical averages of different observables on a large
ensemble of identically prepared systems. In practice, however, we
are usually given with a very limited number of the identical
copies. Any measurement at such limited ensemble provides us with
partial information about the state of the system. This leads to an
important problem of the optimal extraction of information from
finite ensembles of quantum systems \cite{Massar:95,Derka:98}.

\begin{figure}
\begin{center}
 \includegraphics[scale=0.5]{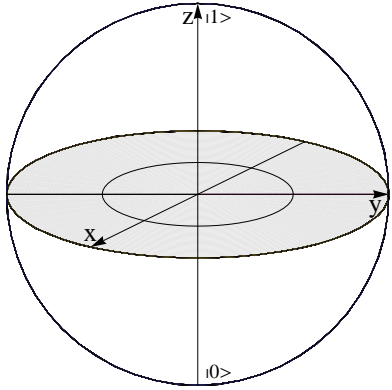}
 \caption{Bloch sphere representation of a qubit state.}
 \label{fig.1}
\end{center}
\end{figure}

In this paper we focus on the best strategy for extraction of
information about fidelity between two finite ensembles of unknown
equatorial qubit states. In order to simplify our discussion we
assume at the moment that each ensemble contains $N$ separable
particles initially prepared in pure states $\ket{\psi_a}$ and
$\ket{\psi_b}$. A pure equatorial qubit state can be parameterized
with a single parameter as
\begin{equation}
 \label{states}
 \ket{\psi_k} = \frac{1}{\sqrt{2}} \left( \ket{0} + e^{i \phi_k} \ket{1}
 \right) \, ,
\end{equation}
where $k$ stands for subindexes $\{a,b\}$ which refer to the
different ensembles and $\{\ket{0}, \ket{1}\}$ is a computational
basis. A pure equatorial state (\ref{states}) can be visualized as a
point lying on a big circle which is formed by the intersection of
the Bloch sphere with $x-y$ plane. All pure equatorial states are
displayed in Fig.~\ref{fig.1} with the big circle.

What is the best strategy to estimate the fidelity
\begin{equation}
 \label{fidelity}
 F_{a,b} \equiv |\spr{\psi_a}{\psi_b}|^2 =
 \frac{1}{4} |1 + e^{i (\phi_b-\phi_a)} |^2
\end{equation}
between the finite ensembles of equatorial qubit states
$\ket{\psi_a}$ and $\ket{\psi_b}$? This question touches the very
foundation of quantum information theory and takes its place among
such widely discussed problems as the state estimation
\cite{Massar:95,Derka:98}, the state discrimination
\cite{Barnett:09,Bergou:10} and the state comparison
\cite{Barnett:03}. Apart of academic interest, moreover, the
fidelity estimation problem may be relevant in implementation of
schemes for quantum communication with linearly polarized photons
and for linear optics quantum computation \cite{Kok:07}. For
example, we are given with a finite ensemble of $2N$ identical
linearly polarized photons. \cor{Each photon in the ensemble can be
described by} some quantum state $\ket{\psi_a}$. \cor{Suppose that}
a half of the photons from the ensemble is subjected independently
to some unitary evolution so that the outputs are in the state
$\ket{\psi_b}$. We like to know the effect of the unitary evolution
by comparing the phases of the states $\ket{\psi_a}$ and
$\ket{\psi_b}$. \cor{This effect can be quantified with the fidelity
(\ref{fidelity})}.

The simplest strategy to estimate the fidelity between the ensembles
of states $\ket{\psi_a}$ and $\ket{\psi_b}$ is to perform state
estimation of each of these states independently on each other and
compute the fidelity (\ref{fidelity}) between the estimated states
$\ket{\psi_a^\prime}$ and $\ket{\psi_b^\prime}$. \cor{Such a
strategy for the fidelity estimation may be called the
measurement-based.} An optimal scheme to estimate the state of
equatorial qubits being given $N$ identical replicas was proposed by
Derka \etal{} \cite{Derka:98}. In this scheme, in more details, a
positive operator valued measurement (POVM), which is characterized
by a set of orthogonal projectors, need to be performed on the
composite system of all $N$ qubits. Since the state of the $N$-qubit
system always remains within the totally symmetric subspace of
$\mathcal{H}_2^{\otimes N}$ where $\mathcal{H}_2$ is the
two-dimensional qubit state space, the dimensionality of the space
in which the POVM need to be defined is $N+1$. If $\ket{n}, \,
n=0,...,N$ is an orthonormal basis in this $N+1$-dimensional space,
the optimal POVM for the state estimation of equatorial qubit is
given by the set of $k=1,...,N$ orthogonal projectors $P_k =
\ket{\Psi_k}\bra{\Psi_k}$ where
\begin{equation}
 \label{povm}
 \ket{\Psi_k} = \frac{1}{\sqrt{N+1}} \sum_{n=0}^N e^{i \frac{2 \pi}{N+1}
 \, k \, n  } \ket{n} \, .
\end{equation}

\begin{figure}
 \begin{center}
 \includegraphics[scale=0.8]{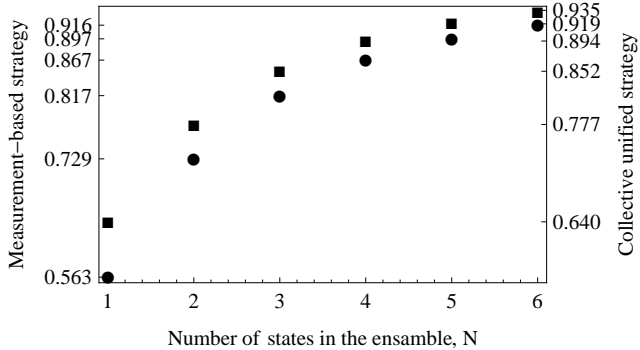}
\caption{The probabilities to reconstruct fidelity (\ref{fidelity})
by the first measurement-based strategy (dots) and the third
collective unified strategy (squares).}
 \label{fig.2}
 \end{center}
\end{figure}

Within this scheme, the maximal mean fidelity between the original
state $\ket{\psi_a}$ and the reconstructed state
$\ket{\psi_a^\prime}$ is given by
\begin{equation}
 \label{mean fidelity}
 \overline{f}(\ket{\psi_a}, \ket{\psi_a^\prime}) = \frac{1}{2} +
 \frac{1}{2^{N+1}} \sum_{i=0}^{N-1} \sqrt{C_i^N \, C_{i+1}^N} \, ,
\end{equation}
where $C_i^N$ and $C_{i+1}^N$ denote the binomial coefficients.
\cor{Later we shall always use term ``probability'' instead of mean
fidelity between estimated and actual values in order to avoid any
confusion. Term ``fidelity'' will be used only with regard to the
value (\ref{fidelity}) of interest.} As indicated above, the scheme
for the state estimation \cor{of equatorial qubits} can be employed
to estimate fidelity between two finite ensembles of equatorial
qubit states. Since the states of interest are estimated
independently, the probability to reconstruct fidelity
(\ref{fidelity}) correctly is given by $\overline{f}^2(\ket{\psi_a},
\ket{\psi_a^\prime})$ and displayed in Fig.~\ref{fig.2} by dots.

An alternative strategy for the fidelity estimation \cor{is based on
quantum cloning and} includes two stages. At the first stage we
provide infinite many copies from available replicas of the unknown
states $\ket{\psi_a}$ and $\ket{\psi_b}$. This task can be realized
with $N \longrightarrow \infty$ equatorial quantum cloning machine
(EQCM) \cite{Bruss:00, Scarani:05}. The copies from the EQCM are
approximate due to fundamental {\it no-cloning principle}
\cite{Wootters:82}. Each copy is given by the mixed state
\begin{eqnarray}
 \label{clones}
 \rho_k^{\rm out} & = & \eta (N, \infty) \ket{\psi_k} \bra{\psi_k}
 + \frac{1}{2}\left[1 - \eta (N, \infty)\right] I \, ,
\end{eqnarray}
where
\begin{equation}
\nonumber
 \eta (N, M) = 2^{M-N} \frac{\sum_{l=0}^{N-1} \sqrt{C_l^N
 C_{l+1}^N}}{\sum_{j=0}^{M-1} \sqrt{C_j^M C_{j+1}^M}}
\end{equation}
denotes the shrinking factor and $I$ is the identity operator.
Having two infinite ensembles of states $\rho_a^{\rm out}$ and
$\rho_b^{\rm out}$ \cor{after the first stage,} one can perform
measurements in some basis and estimate these states by computing
statistical averages. The measurement procedure gives the second
stage of the fidelity estimation. Knowing the estimated states we
can calculate the fidelity (\ref{fidelity}). In the line of this
strategy, the \cor{probability to reconstruct the original state
$\ket{\psi_k}$ from the approximate copies $\rho_k^{\rm out}$} is
given by $f_{EQCM} = \spr{\psi_k | \rho_k^{\rm out}}{\psi_k} =
\left[ 1 + \eta (N, \infty) \right] / 2$. Therefore, the probability
to reconstruct the fidelity (\ref{fidelity}) correctly equals
$f_{EQCM}^2$.

In fact, two discussed strategies for the fidelity estimation are
equivalent in the sense that the probabilities
$\overline{f}^2(\ket{\psi_a}, \ket{\psi_a^\prime})$ and $f_{EQCM}^2$
are equal. This is not surprising and proves the fundamental link
between quantum cloning and state estimation \cite{Bae:06}.

The two strategies above are based on independent estimation of
quantum states and computation of the fidelity using the estimated
states. However, to estimate the fidelity (\ref{fidelity}) we do not
need to know phases $\phi_a$ and $\phi_b$ of the \cor{equatorial
states (\ref{states})}, rather, the difference between them. Based
on this simple observation we now introduce the third two-stage
strategy for the fidelity estimation which unifies previous two
strategies in some sense. At the first stage one takes a pair of
qubits $\ket{\psi (\phi_a)}$ and $\ket{\psi (\phi_b)}$ from
different ensembles and perform a unitary transformation
\begin{equation}
 \label{C-NOT}
\ket{\psi (\phi_a)} \ket{\psi (\phi_b)} \ket{\textbf{A}}_d
\longrightarrow \ket{\psi (\phi_a)} \ket{\psi (\phi_b - \phi_a)}
\ket{\textbf{B}}_d \,
\end{equation}
on these unknown input qubits. Here $\ket{\textbf{A}}_d$ and
$\ket{\textbf{B}}_d$ are states of an auxiliary system before and
after the transformation respectively. The matter of the first stage
is to obtain a qubit in the state $\ket{\psi (\phi_b - \phi_a)}$ at
the output of the transformation (\ref{C-NOT}). At the second stage,
the state estimation of this state is to be performed what allows us
to access \cor{information about the phase $\phi_b - \phi_a$ and by
implication to compute the} fidelity (\ref{fidelity}).

Unfortunately, the transformation (\ref{C-NOT}) can not be performed
exactly on unknown quantum states. This was first pointed out by
Pati and is known today as {\it the general impossibility theorem}
\cite{Pati:02}. Recently, however, an optimal approximation to this
transformation, the so-called universal CNOT gate, was suggested
\cite{Siomau:10}. \cor{This universal gate was obtained by
analytical optimization of a general completely positive map applied
to two pure equatorial qubit states mediated by an auxiliary system.
In construction of the CNOT gate, the standard method for
optimization of cloning transformations was employed
\cite{Scarani:05,Buzek:96,Buzek:98}.} The details about the
construction of the CNOT transformation and its explicit form can be
found in Ref.~\cite{Siomau:10}. For our present discussion, it is
important to note that this universal transformation has similar
structure to the $1 \longrightarrow 2$ EQCM. The output states of
the CNOT transformation are in the mixed states of the form
(\ref{clones}) with $\eta (1, 2)$ and $\ket{\psi_k} = \{ \ket{\psi
(\phi_a)}, \, \ket{\psi (\phi_b - \phi_a)} \}$. \cor{ The
probability to obtain the idealized outputs from the actual output
states of the transformation (\ref{C-NOT}) equals $f_{CNOT} = 1/2 +
1/\sqrt{8}$.}

Repeating the CNOT gate $N$ times on available copies of the states
$\ket{\psi (\phi_a)}$ and $\ket{\psi (\phi_b)}$ we have an ensemble
of $N$ qubits in the mixed state (\ref{clones}) with $\eta (1, 2)$
and $\ket{\psi_k} \equiv \ket{\psi (\phi_b - \phi_a)}$ at the
output. Having this ensemble we can start the second stage -- the
state estimation. In general, state estimation of mixed states with
unknown shrinking factor and phase require construction of a
specific POVM \cite{Bagan:06}. However, in our case the shrinking
factor is known and, therefore, the state estimation of the mixed
state reduces to the estimation of the phase of the pure state
$\ket{\psi (\phi_b - \phi_a)}$. As we discussed earlier, this task
can be accomplished with the POVM (\ref{povm}). Thus the probability
to reconstruct fidelity (\ref{fidelity}) is given by
$\overline{f}(\ket{\psi}, \ket{\psi^\prime}) \times f_{CNOT}$. This
probability is better than the probability to reconstruct fidelity
\cor{within the measurement-based strategy} only for ensembles
consisting of single particle. For ensembles of several particles
the first strategy becomes more efficient.

The reason for the very limited advantage of the third (unified)
strategy over the measurement-based strategy is clear: we applied
the universal transformation (\ref{C-NOT}) only on pairs of qubits
from different ensembles. \cor{A much better strategy} is to apply a
collective transformation on \textit{all} states in two ensembles at
the first stage, i.e.
\begin{eqnarray}
 \label{GC-NOT}
& & \ket{\psi (\phi_a)}^{\otimes N} \ket{\psi (\phi_b)}^{\otimes N}
\ket{\textbf{A}}_d \longrightarrow
\nonumber\\[0.1cm]
   & & \hspace*{2cm}
\ket{\psi (\phi_a)}^{\otimes N} \ket{\psi (\phi_b -
\phi_b)}^{\otimes N} \ket{\textbf{B}}_d \, .
\end{eqnarray}
Due to the general impossibility theorem, this transformation
(\ref{GC-NOT}) can not be accomplished exactly on unknown quantum
states. However, the optimal approximation for this transformation
was recently proposed \cite{Siomau:11}. \cor{The optimal
transformation can be obtained using the same technique for
optimization of a completely positive map as in case of the
transformation (\ref{C-NOT}) and cloning transformations
\cite{Scarani:05}}. The optimal approximation for the transformation
(\ref{GC-NOT}) has similar structure to $N \longrightarrow 2N$ EQCM.
The output states of the transformation (\ref{GC-NOT}) are in the
mixed states of the form (\ref{clones}) with $\eta (N, 2N)$. \cor{
The probability to reconstruct the idealized outputs from the actual
output states is given by $f_{GCNOT} = \left[ 1 + \eta (N, 2N)
\right] / 2$.} Coming to the second stage of the fidelity
estimation, i.e. performing the state estimation on the ensemble of
$N$ output qubits $\rho^{out}(\phi_b - \phi_a)$ with POVM
(\ref{povm}), we obtain that the probability to reconstruct fidelity
(\ref{fidelity}) equals $\overline{f}(\ket{\psi}, \ket{\psi^\prime})
\times f_{GCNOT}$. As displayed in Fig.~\ref{fig.2}, this
probability always superior the probability of the fidelity
estimation by the measurement-based strategy.

\cor{During the discussion of the transformation (\ref{GC-NOT}) we
paid attention only on one of the output ensembles, i.e. on the
ensemble consisting of qubits in the state $\rho^{out}(\phi_b -
\phi_a)$. One may ask how much information about the state of qubits
can be extracted from the other output ensemble of $N$ particles in
the state $\rho^{out} (\phi_a)$. The states of the qubits in this
ensemble can be estimated with POVM (\ref{povm}). The probability to
estimate the state $\rho^{out} (\phi_a)$ is given by
$\overline{f}(\ket{\psi}, \ket{\psi^\prime}) \times f_{GCNOT}$. But,
the state $\ket{\psi (\phi_a)}$ of the initial ensemble of pure
equatorial qubits can be estimated much better with probability
$\overline{f}(\ket{\psi}, \ket{\psi^\prime})$, if the POVM
(\ref{povm}) is applied to this ensemble. Therefore, applying the
transformation (\ref{GC-NOT}) to the given finite ensembles of
equatorial qubits we gain more information about the difference of
pases $\phi_b - \phi_a$ than in case of independent estimation of
these phases and, at the same time, obtain less information about
the single phase $\phi_a$ comparing to state estimation of a single
ensemble. Other words the information that can be extracted  by
measurements about a complex quantum system (consisting of the two
ensembles of equatorial qubits) is somehow conserved independently
on the strategy for the information extraction \cite{Gill:00}.}

At the beginning of the discussion we assumed that both ensembles
contain {\it equal number} of particles in {\it separable} and {\it
pure} states. In fact, the first assumption can be easily remover.
It is easy to define the three strategy for two ensembles with
unequal number of particles $N$ and $K$. By analogy with
transformation (\ref{GC-NOT}), for instance, a generalized $N
\longrightarrow N + K$ transformation can be defined
\cite{Siomau:11}. We checked that the third strategy remains the
best among the three in the case of unequal number of particles in
the ensembles.

However, the other two assumptions are indeed crucial for present
discussion. Being given two ensembles of correlated qubits or qubits
in mixed states, one should accordingly revise all three strategies.
For example, without any knowledge about the shrinking factor of
given mixed states, one should use an optimal set of POVM for state
estimation of unknown mixed states as it was derived by Bagan
\etal{} \cite{Bagan:06}. Moreover, to apply the third strategy on
two ensembles of correlated qubits or qubits in mixed states one
should find an optimal approximation for the transformation
(\ref{GC-NOT}) for such ensembles. It remains an open problem for us
whether the third strategy is still the best in the cases of two
finite ensembles of unknown equatorial correlated qubits or qubits
in mixed states.

\cor{In principle, presented analysis of the fidelity estimation
problem may be repeated for two finite ensembles of arbitrary qubit
states. However, an arbitrary qubit state is characterized by two
phases. This leads to a definition of the fidelity between two
finite ensembles of qubits that includes four phases, i.e twice more
parameters than the fidelity (\ref{fidelity}) for equatorial qubits.
Within the third strategy, moreover, a transformation that allows us
to access the difference between the four phases need to be
constructed. Although the desired transformation has not been
constructed yet, several indications has been obtained that such a
transformation exists.}

In conclusion, we have analyzed the three possible strategies for
the fidelity estimation between two finite ensembles of unknown pure
equatorial qubit states. We showed that the best strategy for the
fidelity estimation includes an optimal universal transformation
(\ref{GC-NOT}) of {\it all} qubits and the state estimation of the
outputs of this transformation by the POVM (\ref{povm}).

It is my great pleasure to thank Stephen M. Barnett for his
encouragement to write this letter and many enjoyable discussions.
This work was supported by the Heidelberg Graduate School for
Fundamental Physics.

\end{document}